\documentclass[sigconf]{acmart}

\usepackage{tabularx}
\usepackage{cleveref}
\usepackage{boxedminipage}
\newcolumntype{Y}{>{\centering\arraybackslash}X}
\newcolumntype{M}{>{\centering\arraybackslash}m}
\newcolumntype{S}{>{\centering\arraybackslash}p{0.085\textwidth}}
\newcolumntype{C}{>{\centering\arraybackslash}c}

\AtBeginDocument{%
  }

\usepackage{todonotes}

\copyrightyear{2025} 
\acmYear{2025} 
\setcopyright{cc}
\setcctype{by}
\acmConference[SIGMOD-Companion '25]{Companion of the 2025 International Conference on Management of Data}{June 22--27, 2025}{Berlin, Germany}
\acmBooktitle{Companion of the 2025 International Conference on Management of Data (SIGMOD-Companion '25), June 22--27, 2025, Berlin, Germany}
\acmDOI{10.1145/3722212.3725635}
\acmISBN{979-8-4007-1564-8/2025/06}

\makeatletter
\gdef\@copyrightpermission{
  \begin{minipage}{0.2\columnwidth}
   \href{https://creativecommons.org/licenses/by/4.0/}{\includegraphics[width=0.90\textwidth]{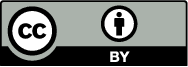}}
  \end{minipage}\hfill
  \begin{minipage}{0.8\columnwidth}
   \href{https://creativecommons.org/licenses/by/4.0/}{This work is licensed under a Creative Commons Attribution International 4.0 License.}
  \end{minipage}
  \vspace{5pt}
}
\makeatother

\begin{document}

\title{Transactional Cloud Applications: Status Quo, Challenges, and Opportunities}
% problems, challenges, ...

\author{Rodrigo Laigner}
\orcid{0000-0003-2771-7477}
\affiliation{%
   \institution{University of Copenhagen}
   \country{Copenhagen, Denmark}
 }
\email{rnl@di.ku.dk}

\author{George Christodoulou}
\orcid{0000-0002-2104-565X}
\affiliation{%
   \institution{Delft University of Technology}
   \country{Delft, Netherlands}
 }
\email{g.c.christodoulou@tudelft.nl}
 
\author{Kyriakos Psarakis}
\orcid{0000-0002-3017-5704}
\affiliation{%
   \institution{Delft University of Technology}
   \country{Delft, Netherlands}
 }
\email{k.psarakis@tudelft.nl}
 
\author{Asterios Katsifodimos}
\orcid{0000-0002-6717-2945}
\affiliation{%
   \institution{Delft University of Technology}
   \country{Delft, Netherlands}
 }
\email{a.katsifodimos@tudelft.nl}

\author{Yongluan Zhou}
\orcid{0000-0002-7578-8117}
\affiliation{%
   \institution{University of Copenhagen}
   \country{Copenhagen, Denmark}
 }
\email{zhou@di.ku.dk}

\renewcommand{\shortauthors}{Rodrigo Laigner, George Christodoulou, Kyriakos Psarakis, Asterios Katsifodimos, \& Yongluan Zhou}

\newcommand{\para}[1]{\vspace{1mm}\noindent\textbf{#1.}}
\newcommand{\parait}[1]{\vspace{1mm}\noindent\textbf{\emph{#1.}}}

\begin{abstract}

% Traditional monolithic applicatins are moving to the cloud as we speak.

Transactional cloud applications such as payment, booking, reservation systems, and complex business workflows are currently being rewritten for deployment in the cloud. This migration to the cloud is happening mainly for reasons of cost and scalability. Over the years, application developers have used different migration approaches, such as microservice frameworks, actors, and stateful dataflow systems.

The migration to the cloud has brought back data management challenges traditionally handled by database management systems. Those challenges include ensuring state consistency, maintaining durability, and managing the application lifecycle. At the same time, the shift to a distributed computing infrastructure introduced new issues, such as message delivery, task scheduling, containerization, and (auto)scaling. 

Although the data management community has made progress in developing analytical and transactional database systems, transactional cloud applications have received little attention in database research. This tutorial aims to highlight recent trends in the area and discusses open research challenges for the data management community.

\end{abstract}

%% The code below is generated by the tool at http://dl.acm.org/ccs.cfm




\maketitle

\section{Introduction}

% \asterios{Why this tutorial:
% Requirements, patterns/architectures, taxonomy, open problems. DB community is missing an opportunity to make a difference: systems folks know nothing about transactions consistency. DB folks need to listen to this. It's a mess ATM. We need a taxonomy and categorization from first principles.}

% \rodrigo{Goal: Emerging software engineering requirements. Both organizational and technical perspectives. Related to advances enabled by cloud. reviewers seem to not understand basic business. technology is at the service of business in order to seek profit and not the other way. claiming organizations should move back to monolithic to solve consistency issues is not a feasible choice economically in most cases.}

% \noindent\textbf{Cloud Completely Changed the Application Development and Deployment Landscape.}

In recent years, many applications such as Customer Relationship Management (CRM), reservation, and payment systems have been migrated to the cloud to take advantage of lower costs and elasticity. These applications were developed as monoliths, typically following the three-tier application architecture (presentation, application, data)~\cite{resurrecting}. In this architecture, business logic is implemented inside the application tier, while all data management takes place in the data tier, typically served by a database management system. 

When migrating such applications to the cloud, developers need to split the functionality of a monolithic application to enable scalable deployment and development efficiency. This approach involves splitting monolithic applications into smaller and independent components that can be deployed and scaled separately, each serving requests as services. This design termed the \textit{microservice} architecture, is widely adopted for migrating applications to the cloud. In the microservice architecture, each microservice is responsible for managing its own data (data encapsulation). Furthermore, implementing complex workflows spanning multiple microservices requires messaging and orchestration.

The emergence of cloud computing as a key paradigm for software and infrastructure as a service has prompted decision-makers and software teams to rethink their strategies for developing, deploying, maintaining, and modernizing their applications. In particular, researchers and industry are currently developing new programming models, data management methods, service communication models, as well as application deployment and lifecycle practices to exploit the low-access barrier to an unprecedented abundance of computing resources provided by the cloud. 
% As computing resources and services in the cloud can be acquired on a pay-per-use basis, that provides greater flexibility and cost savings for users and organizations. 

% At the same time, developers must adapt their applications to the cloud-serving model. To support this transition, specialized deployment systems, such as container orchestrators like K8s and Docker, have emerged to mitigate challenges associated with deploying applications through Virtual Machines (VMs). However, container technologies are oblivious to application state management, just like VMs.

To better align with the goal of on-demand, fine-grained resource provisioning enabled by the cloud and application performance requirements, modular software architectures, such as microservices, distributed application frameworks (e.g., Orleans \cite{bykov_orleans_2011}, Akka Serverless) and the serverless computing paradigm, such as AWS Lambda~\cite{lambda}, emerged as popular alternatives in the cloud application development landscape. 

At the same time, microservices and application runtimes forgo key advantages that monolithic applications have relied on for decades: the delegation of state management, failure recovery, and consistency guarantees to database management systems (DBMS). In modern microservice architectures, these responsibilities are intertwined with the application logic, mixing state management, messaging, and coordination into the application layer. From the perspective of the database community, the situation resembles the early days of computing, when developers relied on ad hoc, application-level transactions to maintain consistency~\cite{papadimitriou1979serializability}.

In the last few years, the database community has focused on building and improving individual components used in cloud applications, such as serverless database systems and stateful functions. However, despite the pressing need for application migration to the cloud, the landscape of runtimes for transactional Cloud applications remains sparse. The programming paradigms and available systems differ substantially, each having key strengths and limitations; a characterization of challenges related to database research is still missing. With this tutorial, we aim to explore the design challenges, clarify key differences between cloud programming paradigms, and highlight open problems and opportunities to evolve the landscape of cloud application runtimes.

\section{Tutorial Overview}
% In this tutorial, we start reviewing the cloud application development landscape, particularly modern software engineering requirements and practices, and how they benefit from resource management, parallelization, and scalability facilitated by the cloud. We then discuss how these cloud-enabled application development and deployment ultimately lead to challenges in maintaining application data consistency.

In this tutorial, we propose a taxonomy (\cref{sec:buildingblocks}) that centers around programming models, state management, and application lifecycle to tame the highly unstructured and heterogeneous cloud application landscape.
The taxonomy reflects the building blocks that practitioners use and sets the stage to explore the state of practice for developing transactional applications in the cloud, along with their different designs and limitations. We then address open issues and research opportunities, highlighting how the database community can play a pivotal role in transforming the landscape of how cloud applications are built in the future.

\vspace{2mm}
\noindent\textbf{Tutorial Outline (3 hours)}
% \vspace{-1.5mm}
\begin{itemize}
    \item \textbf{Context and motivation (45 minutes)}
    \begin{itemize}
        \item Introduction and context
        \item From monoliths to cloud-native applications
    \end{itemize}
    \item \textbf{Building blocks (45 minutes)}
    \begin{itemize}
    \item Programming models 
    \item State management
    \item Messaging
    \end{itemize}
    \item \textbf{Requirements (45 minutes)}
    \begin{itemize}    
    \item Fault-tolerance
    \item Application lifecycle
    \item Scalability
    \end{itemize}
    \item \textbf{Future directions (45 minutes)}
    \begin{itemize}
        \item Open problems 
        \item Research opportunities
    \end{itemize}
\end{itemize}

\vspace{2mm}

\noindent\textbf{Target Audience.} 
The target audience of this tutorial includes PhD students, researchers, and practitioners of different roles (e.g., software and data architects and engineers) who intend to obtain a clear overview of the state-of-the-art cloud application development landscape and its implications for data management.
The tutorial is self-contained and provides the background on scalable cloud applications; no prior knowledge of systems for cloud applications is required. However, familiarity with application architectures helps one understand the materials.

% \george{related work needs to be moved. Not really related work. Everything related should be in the next section.}
% \noindent\textbf{Related Work.}
% Spenger et al.~\cite{survey_actor} survey actor-like programming models for the FaaS paradigm. Laigner et al.~\cite{Laigner2021} characterizes data management in microservices. Our work embodies a larger scope by incorporating multiple paradigms to program data-intensive applications in the cloud. Apart from a wider scope, this tutorial introduces a taxonomy for cloud programming that has not been characterized before. These contributions 
% structure the understanding of the state-of-the-art and facilitate future research.
\section{Building Blocks}\label{sec:buildingblocks}

\begin{figure*}[t]
    \centering
    \includegraphics[width=0.95\textwidth]{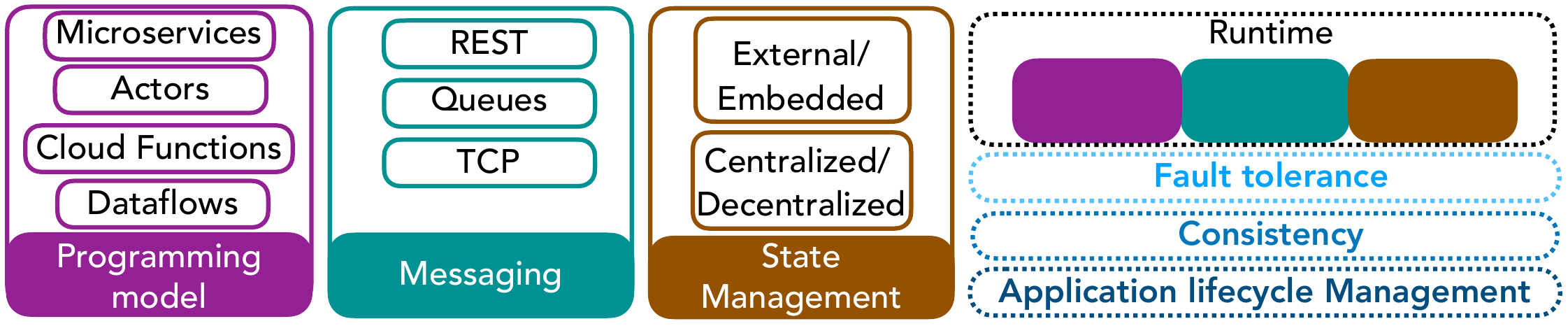}
    \vspace{-3ex}
        \caption{Building blocks and requirements for transactional cloud applications.}
    \label{fig:building-blocks}
    \vspace{-3ex}
\end{figure*}

The programming abstractions offered by systems for developers developing transactional cloud applications are centered around three main building blocks, as shown in \Cref{fig:building-blocks}:
% In this section, as seen in \Cref{fig:building-blocks}, we cover the three building blocks of transactional cloud applications, namely: 
$i)$~programming models ($\S$~\ref{sec:programming-models}), with a focus on their parallelization primitives and how practitioners are building such applications; $ii)$~messaging ($\S$~\ref{sec:messaging}) with focus on different ways of exchanging messages and performing remote procedure calls and finally; $iii)$~state management ($\S$~\ref{sec:state-management}), with a focus on transactions and state consistency across services and scalability.
Their interplay leads to trade-offs concerning programmability, consistency, and performance, as recent findings suggest~\cite{Laigner2021,blanastransactions}.

\subsection{Programming Cloud Applications}
\label{sec:programming-models}
% \kyriakos{merge actors and function-based (with a focus on the stateful functions)}
% \kyriakos{I will remove dataflow since it's not a valid PM}
% \kyriakos{Also, only stateful functions are valid in our case}

Programming models for distributed systems has been a long-standing line of research~\cite{pl_for_distributed, boom, DryadLINQ, hilda, bal1992orca}.
In the context of cloud applications, we identify that programming models play a crucial role in key system aspects, including but not limited to state and message management, fault tolerance, lifecycle management, and scalability.
% In this tutorial, we focus on the programming models used to build scalable cloud applications. 
The status quo is the use of microservice frameworks (e.g., Java Spring~\cite{java_spring}, Python Flask~\cite{python_flask}) and emerging programming models, namely Actors (e.g., Akka~\cite{akka}, Orleans~\cite{bykov_orleans_2011}) and Stateful Functions (e.g., Flink Statefun~\cite{statefun}, Azure Durable Functions~\cite{durable_functions}), all differing significantly in system model, abstractions, and guarantees offered to developers. 

% that mainly aim to streamline the scalability of cloud applications. 

Therefore, we kick off this tutorial by discussing the variables that drive developers to decide on a programming model, namely: $(i)$ the programming paradigm, which relates to the application abstractions exposed to users (e.g., functions, actors, or objects); $(ii)$ code modularity, which includes not only how program modules cooperate but also how application state is partitioned and encapsulated; and $(iii)$ concurrency \& transactional semantics.

Note that this tutorial does not aim to provide in-depth analysis and formal semantics of models; rather, it focuses on how they fit in the cloud landscape and the main limitations of the systems enabling them.

\label{subsub:stateless}

\para{Microservice Frameworks} To reap the benefits of parallel processing and loose coupling, the prevalent approach is functionally partitioning the application logic and state into independent components that communicate with each other via synchronous or asynchronous messages~\cite{Base2008}, called microservices. Microservice frameworks, such as Spring Boot (Java)~\cite{java_spring}, Flask (Python)~\cite{python_flask}, and Dapr (C\#/.NET)~\cite{dapr}, provide tools, libraries, and structures to help developers build microservices. These frameworks often include functionalities like Object-Relational Mapping for database interactions, service communication using REST or message queuing, and retrying features for fault tolerance. Each microservice built with such frameworks often employs a multi-threaded application server. Concurrency control and data consistency management are often provided by the underlying database system used by the component and the configured isolation level.

% argus. barbara liskov
%The traditional way to build scalable systems is by dividing the program into tasks that are run by independent processes. Although providing complete freedom on program expressiveness, developers are faced with challenging implementation efforts: (i) the tasks must be able to progress independently to avoid high synchronization costs; (ii) critical sections must be safeguarded by mutual exclusion primitives, such as via lock management or message-passing designs; (iii) developers may need to reason about resource fairness; (iv) in some cases, developers are responsible for mapping program threads to operating system threads.
% Usually involves direct mapping of user units of execution to OS threads
% https://docs.oracle.com/cd/E19620-01/805-4031/6j3qv1oej/index.html

%To reap the benefits of parallel processing while avoiding concurrency bugs such as data races and possible deadlocks arising from mismanaging locks, it is a common design choice to program applications as \textit{stateless} services. In an stateless program, application state is entirely offloaded to an external database system. Whenever data requires being updated, it is pulled from a DBMS, modified and pushed back, being always transient in memory. % The DBMS is responsible for concurrency control, alleviating developers.
%In this model, data is never updated or maintained via own application data structures. 

% Although not characterized as a programming model, it is worthy noting how developers traditionally utilize the cloud to implement their distributed applications.

\para{The Actor Model} The actor model is a programming model for concurrent and parallel computation in distributed systems~\cite{agha_actors_1986}. An actor models a sequential process that performs transformations on the local state based on incoming messages. Actor systems are formed by a composition of actors, which communicate via asynchronous message-passing. Concurrency in actor systems is achieved by pipelining and dynamic creation of actors~\cite{agha_actors_1986}. Traditional actor systems allow programmers to develop systems using low-level primitives by using actor IDs and prescribing their physical locations.

%\kyriakos{Summarize this and possibly merge with actors}
%\para{Virtual Actors}
Virtual actors~\cite{bykov_orleans_2011} are an extension of the traditional actor model that provides location transparency without forcing developers to deal with actor allocation in a cluster, life-cycle management, explicitly creating and tearing down actor instances, as well as failure transparency. %Similar to the actor model, developers can model their applications into fine-grained virtual actors that encapsulate their own states following a single-thread execution model and communicate by asynchronous messages %hat may have an associated response from the caller grain.
%Each object is identified by its type and an associated key (e.g., UUID), which must be referred when messaging a grain.
It is currently found in popular distributed application frameworks like Orleans~\cite{bykov_orleans_2011} and Dapr~\cite{dapr}.

% Programming abstraction for designing concurrent applications. A programming model where each application-defined entity (i.e., actor) is uniquely identified, encapsulates its own private state (no memory sharing) and communicate with each other via asynchronous message passing. Messages are processed sequentially and each spawn a mapped operation (often operating over the private state).

\para{Cloud Functions} With the emergence of serverless computing~\cite{serverless}, a new cloud paradigm called Function-as-a-Service (FaaS)~\cite{bench_faas,durable_functions} rose in popularity. In FaaS, developers build applications as a collection of functions. Function executions are triggered by external events, such as clicks or invocations from other functions, allowing for function workflow compositions.

% However, FaaS did not provide state management primitives or consistency and fault tolerance guarantees and could not be used to create cloud applications. 
Initially, FaaS offerings targeted workloads with small to moderate I/O and communication, demotivating offering data models and consistency guarantees on operations within a single function or cutting across functions~\cite{beldi,cloudburst}.
More recently, though, there has been increasing interest in extending the FaaS paradigm to applications that access state intensively, called Stateful-FaaS (SFaaS)~\cite{beldi, boki, cloudburst, tstatefun, styx}. In SFaaS, developers also write programs based on composing functions and enjoy a key-value interface to access the global application state. Apart from the shared state interface, the programming, execution, and deployment model resembles Virtual Actors.

\para{Stateful Dataflows} The dataflow model prescribes that an application is represented as a data flow graph. That involves decomposing programs into independent processing units. Organized as Directed Acyclic Graphs (DAGs), processing units (nodes) exchange data via message streams (edges).
% A processing unit cannot determine the order of streams from different sources without a merge operator.
Dataflows have been mainly applied as the programming model for analytical batch and stream processing systems like Flink~\cite{flink}. In these systems, processing units are framed as operators that can perform either stateful (e.g., joins, aggregates) or stateless (e.g., map, filter) operations. Message streams can be partitioned and assigned to different operator instances that execute concurrently. Stateful operators typically do not share state, preventing concurrency issues and enhancing parallelism.

However, the dataflow model has two main issues regarding its use for transactional cloud applications. First, dataflow systems are typically programmed using functional programming-style dataflow APIs, requiring developers to rewrite cloud applications to align with the event-driven dataflow model. While many cloud applications can be adapted to this paradigm, doing so demands significant programmer training and effort. Second, implementing \textit{transactions} on top of dataflows, namely transactions that span multiple services with serializable guarantees, is still an open problem~\cite{statefun, zhang2024survey, DEHEUS2022102015, styxcidr}.

\subsection{Messaging}
\label{sec:messaging}

For any two components of a cloud application to communicate, a form of remote procedure call (RPC) is required, i.e., a way for a component to call a function on another remote component. In the past, RPC has taken multiple forms, such as Java's RMI~\cite{java_rmi} or CORBA's OMG~\cite{vinoski1997corba}. Nowadays, most applications opt for either using REST APIs or message queues. We detail those below. 

\para{REST and gRPC}
Built over HTTP (or HTTP/2), REST and gRPC are among the most popular ways to implement remote procedure calls for messages exchanging in microservice architectures.
%A web server receives a HTTP call with an encapsulated REST message and the web server relays the message to an applciation server (e.g., a microservice or an actor). 
HTTP-based protocols~\cite{indrasiri2020grpc} are typically stateless and cannot provide guarantees of message delivery. 
Thus, applications requiring message delivery guarantees must ensure these at the application level. Independently of the HTTP protocol adopted, a unique ID (e.g., in the form of an \texttt{idempotency key}~\cite{idempotence}) is traditionally leveraged to prevent the execution of non-idempotent operations for incoming duplicated messages. Messages are often duplicated in two cases: partial failures in the sender side and redelivery after a timeout. However, uniqueness ID guarantee and subsequent detection of duplicated messages are still the responsibility of applications~\cite{deduplication},
% used to deduplicate message delivery retries at the receiver side. \texttt{Idempotency-key}s are managed by the developers in application-level code, 
adding to the complexity of developing cloud applications. 

\para{Message Queues}
Message queues (e.g., Apache Kafka~\cite{kafka}, RabbitMQ~\cite{rabbitmq}, RedPanda~\cite{redpanda}, etc.) are typically used to implement asynchronous applications. The sender first pushes a message into a queue, the queue persists the message, and the receivers asynchronously pull messages from the queue to consume them. Producers and consumers are decoupled in time, facilitating the support to partial failures~\cite{tanenbaum2008distributed}. On the other hand, receivers require acknowledging the consumption of messages. Despite the apparent simplicity, applications must coordinate the message processing and subsequent acknowledgment to prevent the execution of non-idempotent operations, a challenging task for many developers~\cite{Laigner2021}.
% When message queues are used, deduplication becomes relatively easier, as the sender does not need to explicitly re-send a message, and the receiver can inform the message queue when it has finished processing a message. 

\para{Relation of Messaging \& State} 
As described above, an application state mutation depends causally on the arrival of a message. The operations over the state resulting from an incoming message must reflect in the receiver's state exactly once, characterizing the exactly-once processing guarantee.
In sum, this means that the sender should be able to re-send messages to ensure the receiver has received them and, if a message is received multiple times, the receiver should be able to deduplicate them. % receipt of a set of messages, and 

% \highlight{2}{Implementing scalable applications via microservices leads to challenges in ensuring message delivery and processing guarantees. Although resorting to programming models for distributed applications alleviates some of these challenges, it forces developers to adapt their application code to non-conventional paradigms. \textbf{can delete if necessary, just an after thought...}}

\subsection{State Management}
\label{sec:state-management}

The state of an application usually refers to an application's data (e.g., the contents of a shopping cart or a bank account) that impacts its functionality and responses to client requests. Orthogonally to programming models and messaging, state management involves the placement and movement of data across components and strategies to make data durable and consistent. As depicted in \Cref{fig:building-blocks}, state management in cloud applications depends on two main decisions: $i)$ whether the state will be managed using an \textit{embedded} approach (residing within the application runtime) or an \textit{external} system, such as a database or a blob store, and $ii)$ if the state access will be \textit{centralized} or \textit{decentralized}. In a centralized approach, the system manages the whole state in a unified way. In a decentralized approach, every subcomponent (e.g., individual operators in a stream processing system) handles its state independently. In this section, we discuss the state management design space in cloud applications.

\para{Microservices} There are two approaches to manage state in microservice architectures~\cite{Laigner2021}: 
$i)$~shared database, where data is logically separated (e.g., through private tables or distinct schemas), sharing database resources (i.e., centralized database); and $ii)$~database per service (i.e., decentralized), where each service enjoys a dedicated database server, ensuring physical data isolation. 

Physical isolation offers reduced coupling and independent scalability at the expense of higher complexity and infrastructure costs. On the other hand, a physically centralized database can impact teams by sharing database resources and artifacts (e.g., memory and disk resources, locks, or latches), jeopardizing performance isolation and application upgrades, respectively. Note that microservice architectures typically opt for the \textit{external} approach to data management ($\S$~\ref{fig:building-blocks}). 

\para{Actors}
Actor systems enforce logical state isolation, i.e., each actor manages and mutates its state. To this end, they often leverage language runtime and framework support to decouple actor calls from the actual execution (i.e., actor instantiation and memory address), inhibiting users from dealing with resource-sharing concerns~\cite{Liu2024}. Although actors typically keep their state private in main memory, some actor frameworks offer state management APIs that allow developers to store memory-resident states in durable storage~\cite{orleans_best_practices}.
% If every piece of software that handles different requests is separated, there is no contention, leading to faster execution. 
Updates to an actor's state are only possible by messaging the actor. 
Depending on the actor system, mapping actors to servers can be both manual or dynamic ~\cite{bykov_orleans_2011}. Although it does not often affect how users operate over the actor state, actor placement can impact performance.
In any case, data freshness guarantees are tied up to the most recent actor communication.

\para{Cloud Functions} There are two dominant models for state management in the FaaS paradigm: private or shared state~\cite{cloudburst}.
While in the former, the state of a function is modeled as an object that is tied to a given function, in the latter, functions are free to access any object, subject to the concurrency model imposed by the FaaS execution platform.
% In both cases, whether the FaaS platform allows for interleaving function invocations (e.g., scheduling function code while a previous invocation of the same function is waiting for an I/O operation), functions often execute as a single-thread model.

% Compared to abstraction that embeds together state and behavior, such as \textit{virtual actors} in Orleans and \textit{entities} in Statefun, in FaaS, developers write programs based on composing ephemeral functions that are uniquely identified~\cite{serverless}. At the same time it allows specifying programs at a finer granularity, it can increase the complexity of developing highly-interactive, complex object-oriented programs.

FaaS systems often ensure function invocations are scheduled in an individual computational resource, such as a container or a virtual machine~\cite{serverless}. In the first case, whenever a function is triggered, the corresponding state is brought from disaggregated storage to the memory of the compute nodes assigned to run the respective function, all transparent to user code. In the second case, operations on shared state necessarily incur network round trips.

\para{Dataflows} In most distributed dataflow systems, the application state is decentralized by design~\cite{flink}. Typically, operators are scheduled for execution in separate nodes and rely on embedded LSM-based key-value stores like RocksDB~\cite{rocksdb} as a local state. Whenever the operator's state exceeds the local storage capacity, the state must be checkpointed, and the associated operator must be migrated to another node with sufficient storage capacity. Recently, there has been increasing interest in using tiered storage to battle scenarios where operators' states exceed local node storage~\cite{risingwave,flink2}. In this case, cloud object storage systems like S3 are used not only for checkpointing states~\cite{flink} but also to store operators' states. 
It is worth noting that, unlike service-based architectures, state management in dataflow systems is transparent to developers.
% not so much in upgrades though...
% That requires optimizations on wirtes, like buffered writes, and holistic management of data in memory and storage, a non-trivial task.

% Although systems offer sharing state capabilities, data should stay immutable and accessed in a read-only fashion. 

\subsection{Discussion} 
It is entirely possible to have a combination of programming models and state management primitives. For instance, Orleans can make use of an external database to store actor state, while there are dataflow engines that may store their state, instead of internally, to an external database system~\cite{gustafson2022}. Although these approaches depart from the strict limits of the programming model or architecture at hand, they are valid deployment scenarios that are used in practice. 
In addition, low-latency microservices may need to embed a state to enhance data locality. Typically, a \textit{cache} (e.g., Redis or Hazelcast IMDG) is used to speed up state retrieval, blurring the line between embedded and external state management. In any case, while mixing and matching different systems and approaches, deployments that go beyond the traditional settings also come with consequences in terms of fault tolerance and scalability, which will be discussed in the next section.

\section{Requirements}
\label{sec:requirements}
In this section, we discuss the cross-cutting requirements of transactional cloud applications, namely: $i)$~fault-tolerance ($\S$~\ref{subsec:fault}); $ii)$~consistency ($\S$~\ref{subsec:consistency}) and $iii)$~application lifecycle management ($\S$~\ref{subsec:lifecycle-management}). % with a focus on code updates, (auto)scaling, and state migration. 

\subsection{Fault-tolerance}
\label{subsec:fault}
\para{Microservices}
Fault tolerance in microservices is achieved by making the application logic stateless and leaving state handling to an external database. Therefore, as long as a database of a given service is alive, the service operates normally. In case of failure at the stateless (application logic) microservices side, it is enough to restart a new service and connect to the same database. Although fault-tolerant by design, microservices may pose issues concerning state \textit{consistency} due to the lack of a strong message delivery guarantee or transactional guarantee for multi-service workflows.  

\para{Actors} Modern actor systems have traditionally empowered three-tier architectures~\cite{bykov_orleans_2011,Liu2024}, so developers checkpoint actor states to an external DBMS to ensure durability. As actor frameworks do not impose a database deployment model, ensuring performance, access, and failure isolation at the database tier is a non-trivial task at the hands of developers. On the other hand, actor frameworks like Orleans offer failure transparency by migrating actors across nodes in the presence of partial failures~\cite{bykov_orleans_2011}. However, weak message delivery semantics and lack of transactional guarantees can leave actor states inconsistent after failures ($\S$~\ref{subsec:consistency}).
% necessitaitng manual intervention

\para{Stateful Dataflows} For recovery, dataflow systems rely on checkpointing and logging mechanisms. Checkpoints in a distributed environment can be either independent per worker or in coordination by using a protocol \cite{ChandyL85}. Checkpointing ensures that the entire state is saved in (external) durable storage, and logging keeps track of all the data accesses between checkpoints. On failure, the system can retrieve its state by reloading the latest checkpoint, recalculating the state based on the actions saved in the log, and continuing from where it was left off.

% \rodrigo{I started the text below, but missing creativity now. maybe we should add a little bit more about SFaaS?}
% SFaaS systems like Statefun, by relying on Apache Flink processing model, enjoys transparent recovery of application objects and messages.

\subsection{Consistency}
\label{subsec:consistency}
The consistency models in distributed systems reason about reads and writes on shared state and their real-time order guarantees across processes~\cite{tanenbaum2008distributed}.
In cloud programming paradigms, though, we observe that the consistency models are inherently driven by the enabler systems' communication model and state management properties.
%In the last section, we defined a programming model for the cloud as the combination of a programming paradigm and modularization properties, including the data partitioning scheme across modules, the concurrency semantics of module executions, and how modules exchange messages.
% Programming models for the cloud bound the set of possible histories through the data partitioning and the inherent concurrency semantics ($\S$~\ref{subsec:prog_model}). \textcolor{red}{On the other hand, 
% they introduce an additional set of operations that, although not data items, are part of the state and key for transactional semantics, the so-called asynchronous messages.} In case of never arriving, reprocessing, and out-of-order processing, the system's partial order of operations is affected. This section discusses the role of messages in consistency models for cloud applications in weak and strong models.

\para{Microservices}
Distributed applications designed through microservice architectures often remount the idea of the BASE model~\cite{Base2008}, characterized by eventually consistent application partitions through queuing operations~\cite{life_beyond}. Practitioners also refer to this eventual consistency model through sagas~\cite{sagas} or patterns like orchestration and workflows~\cite{OrchestratingServerless}. 

Microservices often avoid distributed commit protocols to decouple components~\cite{Laigner2021}.
% In order to ensure transactional guarantees in service-oriented architectures, developers must use distributed commit protocols, such as 2-Phase Commit (2PC). 
That would involve using language-specific libraries and implementing the protocol phases in each microservice, a complex and error-prone task for general developers.
Besides, enabling the protocol across services is often impossible due to the lack of library support across heterogeneous programming languages and databases~\cite{epoxy}. Most importantly, directly accessing data items in external services may break the desired state encapsulation, while the blocking nature of traditional protocol implementations affects performance.

\para{Actors}
With at-most-once messaging delivery guarantees by default, weak consistency across components is a popular design choice in actor-based applications. Some actor systems like Orleans allow customizable timeouts for retries to achieve at-least-once delivery. Statefun differs from Orleans in managing state updates and messages in an integrated manner, transparently rewinding the application state to a previously consistent checkpoint in case of a delivery error. Therefore, it achieves exactly-once processing and atomicity as a consequence. However, there is no transactional isolation across Statefun entities. % By default, Orleans provides no transactional isolation across actors. 

To enable transactional serializability in Orleans, users must utilize the Transactions API~\cite{OrleansTxnDoc}. Apart necessitating porting the actor attributes to opaque objects~\cite{Liu2024}, it has been shown to introduce a significant performance penalty according to recent experimental evaluations~\cite{snapper,Laigner2024}, demotivating broader adoption.

\para{Cloud Functions}
Cloudburst enriches functions with causally consistent shared state accesses through a key-value abstraction~\cite{cloudburst}.
% https://angelhof.github.io/files/papers/netherite-2022-vldb.pdf
Durable functions~\cite{netherite}, in the context of Azure Durable Functions service, enhance FaaS with the ability to model entities (i.e., typed objects) as state abstractions for function manipulation, a richer state management abstraction than traditional FaaS offerings. Furthermore, individual function operations are atomic and enjoy exactly-once guarantees, guaranteeing atomicity in function compositions. Users must acquire and release locks explicitly to ensure transactional isolation on operations involving multiple entities (e.g., transfer money between account entities). However, there is no support for transactional isolation across functions.

Another category of Cloud Function systems goes beyond by providing transactional serializability on computations cutting across functions~\cite{beldi,boki}. However, recent work~\cite{styx} has found challenges in supporting large-scale, complex transactional applications like TPC-C in existing state-of-the-art SFaaS systems.

\para{Stateful Dataflows}
The dataflow programming model rose in popularity due to stream processing engines with support to exactly-once processing guarantees~\cite{flink, ZahariaDLHSS13}. Exactly-once guarantees eliminates the need for fault-tolerant code in the application since the engine transparently handles failures. However, exactly-once processing guarantees alone cannot ensure transactional isolation.

\subsection{Application Lifecycle Management}
\label{subsec:lifecycle-management}

\para{Resource Management}
The programming abstractions offered to developers ($\S$~\ref{sec:programming-models}) also play a key role in application lifecycle management. In microservice frameworks, application maintainers are in charge of deploying services, detecting failures, and implementing recovery routines. These implicitly include the objects managed by the application at run-time, complexities that only exacerbate the existing challenges of maintaining consistent application states ($\S$~\ref{sec:state-management}). These challenges motivated the development of distributed frameworks that transparently manage the lifecycle of applications. 
% In frameworks that expose virtual actor abstraction, such as Orleans, users enjoy location and lifecycle transparency. Orleans allocates virtual actors on demand in healthy computational resources and deallocates resources used by actors when they are no longer used. In case of failures, Orleans transparently migrates virtual actors, ensuring the application remains functional. However, users must still handle resource provisioning and scaling explicitly.
Serverless computing and FaaS offer transparent resource provisioning, function scheduling, failure handling, and elasticity to application maintainers. However, challenges associated with cold starts, execution performance, and costs undermine a wider adoption of the FaaS paradigm in application architectures~\cite{serverless}.

\para{Application Evolution}
The evolution of applications is a key concern in the software engineering lifecycle, and it is no different in cloud applications~\cite{se_cloud,yau2011software}. In a distributed environment, this includes, but is not limited to, the deployment, upgrading, and deprecation of components, as well as changes in the data and event schema. Surprisingly, support for application evolution in cloud applications is limited, and upgrades are often handled via ad-hoc approaches that rely on the expertise of application maintainers for correctness. In this tutorial, we cover the application evolution space for service-oriented architectures, actors, and dataflow systems.

\section{Open Problems \& Research Opportunities}
In this section, we describe a set of open problems in programming models ($\S$~\ref{sec:models-systems-open}), state and messaging ($\S$~\ref{sec:state-messaging-open}) and benchmarks ($\S$~\ref{sec:benchmarks-open}).

\subsection{Programming Models \& Systems}
\label{sec:models-systems-open}

The variety of programming models available, along with the associated trade-offs in designing applications, such as data partitioning, access, and storage, concurrent application logic execution, fault handling, upgrade support, guarantees during crashes and network partitions, pose challenges to application developers in deciding for an ideal model. Another factor that only exacerbates these challenges is the proliferation of terms like 
\textit{entities} or \textit{objects}~\cite{statefun,OrchestratingServerless,service_fabric}, \textit{workflows}~\cite{dapr,OrchestratingServerless}, \textit{durable}~\cite{OrchestratingServerless}, \textit{stateful}~\cite{statefun,OrchestratingServerless,service_fabric, styx}, \textit{reliable}~\cite{service_fabric}, and \textit{virtual}~\cite{bykov_orleans_2011,dapr}, which are not consistent across systems since they express varied guarantees for applications.
% many concepts are duplicated, intertwined, which can be challenging for newcomers and designing complex applications.
% formal model facilitate understanding ad implementation.

Apart from the dataflow and actor models~\cite{agha_actors_1986}, and more recently Durable Functions~\cite{durable_functions},
many programming models used in the cloud today are not formalized. The lack of formalizations and semantics of programming models hinders the ability to reason about a cloud application's desirable properties (such as safety, liveness, and consistency), 
a key impediment to advancing cloud programming. 
Another direction that can mitigate some of these concerns is via declarative programming. Ongoing work in this realm includes stateful entities~\cite{psarakis2021stateful}, HydroLogic~\cite{directions_cloud}, and event-based constraints~\cite{event_constraints}.

Furthermore, systems should be designed to enable developers to effectively perform traditional software engineering activities. Limitations with debugging, application evolution, and observability are additional factors that demotivate the use of systems for cloud programming.
An open question is whether it is possible to devise a programming model and system with transparent parallelization, scalability, and consistency.

\subsection{State \& Messaging}
\label{sec:state-messaging-open}

\para{Data Model}
Programming models used in the cloud often provide opaque state management abstractions~\cite{Liu2024,bykov_orleans_2011,psarakis2021stateful,DBLP:conf/cidr/Helland07}. To fully realize the benefits of serverless computing, it is key that programming abstractions for the cloud offer not only formal semantics but evolve to allow users to operate with richer data models and express cross-component data invariants \cite{directions_cloud} that are popular in practice~\cite{Laigner2024}, not jeopardizing state encapsulation.

\para{Disaggregation}
In the same line of industry-strength cloud-native database~\cite{aurora} and stream processing systems~\cite{jet,flink2}, disaggregated storage~\cite{DisaggregatedDb} can be leveraged in cloud programming abstractions to support ever-growing data that applications process in the cloud. That must account for performance, access, and failure isolation properties, abstracting away these concerns from application code. Although most FaaS architectures provide disaggregated storage by design~\cite{serverless}, challenges inherent to composing applications via ephemeral functions and shared state access, hindering programmability and encapsulation of states of cloud applications, respectively, are still present.
% Although useful for many I/O-intensive tasks and use cases on which concurrent access to data items is less frequent, such as ML models, % In any case, by abstracting away resource management, task scheduling, lifecycle management from users, a serverless cloud model often disaggregates compute and storage. 

Most microservice architectures adopt persistent and asynchronous communication via message queue systems~\cite{tanenbaum2008distributed}, with the message layer being disaggregated by design. However, considering most cloud applications rely on language runtimes such as Java's JVM and .NET's CLR, optimizing runtimes for cloud applications could focus on optimizing message transport, processing, storage, and recovery, considering the operating system and application interplay.
% in transactional cloud applications with decoupled message layer.
% Another line that has been little interest, but not less important is message storage. 
% what about event management? they can be entirely in main memory or made durable. affects performance. usually not addressed by database researchers. also trend on tiered storage for kafka. 
% independently of the model, disaggregation is a key concern at scale.

\para{Consistency}
Traditional approaches such as SAGAs \cite{sagas} and OpenXA \cite{specification1991distributed} allow for coordinating consistency guarantees across microservices. More recent work~\cite{antipode} introduces causal consistency for microservice architectures. Cross-engine transactions~\cite{cross_engine} is a promising approach since it operates at a lower level than the application. However, implementations should avoid exposing private encapsulated data and protocol details in application code. Coordinating with external, often legacy, systems is very common in cloud applications that developers currently handle in an ad-hoc fashion.

\subsection{Workloads \& Benchmarks}
\label{sec:benchmarks-open}

Benchmarking a distributed cloud application for performance and even correctness is largely a task that takes place in an ad-hoc fashion at the moment. Efforts such as DeathStar~\cite{deathstar} have been used to evaluate distributed cloud application frameworks~\cite{beldi,boki} alongside TPC-C~\cite{styx}. In addition, despite recent efforts to benchmark cloud applications~\cite{deathstar,train_ticket,online_boutique}, most benchmarks are oblivious to key aspects of data management. At the same time, traditional metrics such as throughput and latency used to benchmark OLTP and OLAP systems may not suffice emerging cloud programming systems alone. Modeling request arrivals should consider systems' design goals and the cloud serving model used~\cite{schroeder2006closed}. 

The use of event streams as a paradigm to compose applications and the presence of data invariants, transactional guarantees, data replication, and querying in real-world applications are examples of missing requirements for existing benchmarks. Recent work~\cite{Laigner2024} aims to fill these gaps, but challenges related to dynamic workloads, observability, and recovery remain open.

\section{Biographies}

\noindent\textbf{Rodrigo} is a PhD Fellow at the University of Copenhagen. His research lies on devising effective programming abstractions and efficient systems for emerging data-intensive applications. During his doctoral studies, he published relevant articles about distributed data-intensive applications.

\noindent\textbf{George} is a Postdoctoral Researcher at TU Delft. His research centers around indexing, as well as scalable and efficient data management with a particular emphasis on stream processing, and distributed systems.

\noindent\textbf{Kyriakos} is a PhD candidate at TU Delft building systems for scalable cloud applications. This includes Styx, a deterministic transactional dataflow system that offers a Stateful-FaaS API for creating scalable cloud applications.

\noindent\textbf{Asterios} is an Asst. Professor at TU Delft, working on scalable data management, focusing on cloud application runtimes, stream processing, and data integration. Asterios is one of the receivers of the ACM SIGMOD Systems award in 2023. 

\noindent\textbf{Yongluan} is a Professor at the University of Copenhagen. His research interests span database and distributed systems, with his recent focus on scalable event-driven systems. He has authored over 80 peer-reviewed research articles in international journals and conference proceedings. 

\begin{acks}
This work was partially supported by Independent Research Fund Denmark under Grant 9041-00368B, as well as the Vidi research program project number 19708, financed by the Dutch Research Council (NWO).
\end{acks}

\balance

% \newpage

\bibliographystyle{ACM-Reference-Format}
\balance
\bibliography{main}

% \newpage
% \appendix
% \input{sections/06_appendix}

\end{document}